\documentclass[12 pt]{article}
\usepackage{times,bm}
\usepackage[affil-it,]{authblk}
\usepackage[pagebackref=false,colorlinks=true,linkcolor=blue,citecolor=blue,urlcolor=blue]{hyperref}
\usepackage{geometry}
\usepackage{graphicx}
\usepackage{cite}
\usepackage{amsmath}
\usepackage{amsfonts}
\usepackage{soul}
\usepackage{amssymb}
\usepackage{subfigure}
\usepackage{color}
\usepackage{float}
\usepackage{multirow,multicol}
\usepackage{tabulary}
\usepackage[flushleft]{threeparttable}
\usepackage{pgfplots,subfigure}
\usepackage{xcolor}
\numberwithin{equation}{section}
\usepackage{tikz}
\usepackage{placeins}
\usepackage{caption}
\usepackage[normalem]{ulem}
\usepackage{hyperref}
\usepackage{nameref}
\usepackage{comment}
\usepackage{pgfplots}
\pgfplotsset{compat=1.7}

\usepgflibrary{arrows}
\usetikzlibrary{shapes.callouts}
\tikzset{
	level/.style   = { thick, },
	connect/.style = { dotted, red   },
	notice/.style  = { draw, rectangle callout, callout relative pointer={#1} },
	label/.style   = { text width=2cm }
}

\usepackage{letltxmacro}
\makeatletter
\let\oldr@@t\r@@t
\def\r@@t#1#2{%
	\setbox0=\hbox{$\oldr@@t#1{#2\,}$}\dimen0=\ht0
	\advance\dimen0-0.2\ht0
	\setbox2=\hbox{\vrule height\ht0 depth -\dimen0}%
	{\box0\lower0.4pt\box2}}
\LetLtxMacro{\oldsqrt}{\sqrt}
\renewcommand*{\sqrt}[2][\ ]{\oldsqrt[#1]{#2}}
\makeatother

\begin{document}
	\newcommand{{\ri}}{{\rm{i}}}
	\newcommand{{\Psibar}}{{\bar{\Psi}}}
	\newcommand{{\red}}{\color{red}}
	\newcommand{{\blue}}{\color{blue}}
	\newcommand{{\green}}{\color{green}}
	\newcommand{\rev}[1]{\textbf{\textcolor{red}{#1}}}
	
	\title{On Geodesic motion of particles with zero energy in Kerr and rotating dirty black holes}

	\author{\large  
        \textit{V. Vertogradov\footnote{E-mail: vdvertogradov@gmail.com}$^{\ 1,2}$}, \textit{L. Shleiger}$^{\ 3}$\\
        \small \textit {$^{\ 1}$SPB branch of SAO RAS, 65 Pulkovskoe Rd, Saint Petersburg 196140, Russia.}\\
       \small \textit {$^{\ 2}$Physics department, Herzen state Pedagogical University of Russia, 48 Moika Emb., Saint-Petersburg 191186, Russia.}\\
        \small \textit {$^{\ 3}$Ioffe Institute, 26, Politekhnicheskaya Str., 194021, Saint-Petersburg, Russia.}}  
	
	\date{\today}
	\maketitle
	
	\tableofcontents
	\begin{abstract}
		In this work, we explore the properties of timelike geodesic for particles with zero energy. We found some similarities between geodesic motion in Schwarzschild black hole and timelike geodesics for particles with zero energy. We show, that general relativistic corrections might disappear for negative energies. The existence of circular orbits for zero-energy particles might lead to an unbound center of mass energy of two colliding particles. However, in this work, we explicitly show that closed orbits are absent for zero energy particles in generic axially symmetric black holes. We also investigate timelike motion in generic axially symmetric black holes.

	\end{abstract}
	
	\begin{small}
	\end{small}
	
	\FloatBarrier
	
	
	\section{Introduction}

Black holes play an important role in modern theoretical physics and astrophysics. Their surroundings can serve as a super collider of a particle~\cite{bib:grib_complex}. Recent observations by Event Horizon Telescope Collaboration revealed an image of a black hole in the center of M87~\cite{bib:m87} and Milky Way~\cite{bib:way} galaxies. A thorough analysis of its shadow~\cite{bib:docu} showed that these black holes should be rotating ones. Kerr spacetime is one of the simplest metrics that can be used to describe the exterior geometry of a rotating black hole (see~\cite{bib:kerr_review} For comprehensive review). Any black hole in interaction with the nonzero classical matter fields will be referred to as dirty one~\cite{bib:wisser}. Extra matter field can cause changes in exterior geometry and its influence on black hole shadow has been investigated in papers~\cite{bib:ali2024podu} for spherically-symmetric and~\cite{bib:ali2024new} for axially-symmetric black holes.

In the paper ~\cite{bib:bsw}, authors showed that the center of mass energy of two colliding particles might be infinitely large in the extremal Kerr black hole. This effect demands fine-tuning settings of parameters and it was investigated in a series of papers for rotating~\cite{bib:grib_on, bib:santos, bib:joshi_naked, bib:joshi_naked2, bib:zaslav_gen, bib:zaslav_dirty, bib:pavlov_high, bib:japan} and charged~\cite{bib:zaslav_anti, bib:zaslav_charged, bib:zaslav4, bib:vertogradov2024gc} black holes. However, this effect is absent in Schwarzschild spacetime but its analogue is discussed for Vaidya ~\cite{bib:vertogradov2020universe} and Schwarzschild~\cite{bib:pavlov20xxgc} black holes. The collision should occur in the event horizon vicinity with luminous relative speed. This might happen if there are closed orbits near the event horizon or if the particle can have a turning point in the vicinity of the horizon. The nearest unstable closed orbit in Schwarzschild spacetime is a photon rink at $r=3M$ which is too far from this effect. The existence of the turning point demands that an effective potential of timelike geodesics $V_{eff}(r_{tp})=0$. However, an effective potential in general spherically-symmetric spacetime
\begin{equation}
ds^2=-f(r)dt^2+f^{-1}(r)dr^2+r^2d\Omega^2,
\end{equation}
is given by
\begin{equation}
V_{eff}=f(r)\left(\frac{L^2}{r^2}+1\right)-E^2.
\end{equation}
Here $L$ and $E$ are angular momentum-per-mass and energy-per-mass respectively. However, the condition $r_{tp}=r_h$ demands $E=0$, where $r_h$ is the event horizon location. In spherically symmetric spacetime, zero energy particles are absent outside the event horizon.

The situation changes when one considers the rotating black hole. In the rotating case, there exists a Killing horizon outside the event horizon which is called the ergosphere. 
Killing vector which is the manifestation of time-translation invariance $\frac{\partial}{\partial t}$ becomes null on this horizon. In particular, it means that the region between the event horizon and ergosphere, known as ergoregion, becomes spacelike. This fact can lead to the Penrose process~\cite{bib:pen}. According to this process, particles with negative energy can exist in the ergoregion of a rotating black hole due to decay or collision. Penrose process has been investigated in a series of papers for rotating~\cite{bib:pavlov2015mpla, bib:vertogradov2015gc, bib:on_pen, bib:col, bib:is, bib:on_negative, bib:thecol, bib:energy, bib:infty, bib:super}, charged~\cite{bib:richartz, bib:zaslav_negative_charged, bib:rufini, bib:confin} and dynamical~\cite{bib:vertogradov2023ctp, bib:vertogradov2020universe}. Later, it was understood that in the ergoregion of a rotating black hole, there might be particles with zero energy but not zero momentum~\cite{bib:pavlov_zero, bib:pavlov_cosmology}. 

Timelike geodesics in the spherically-symmetric black hole might contain Newtonian attractive force, repulsive centrifugal force, or general relativity corrections which are responsible for anomalous perihelion ~\cite{bib:chandrasekar,bib:vertogradov2023prd}. Timelike geodesics in dynamical spherically-symmetric black holes might contain an additional term that corresponds to induced acceleration ~\cite{bib:misner,bib:yaughoub2017epjc,bib:yaughoub2018epjc,bib:vertogradov2023mpla}. The common feature of these acceleration terms is that they do not depend on the energy of a particle. The situation dramatically changes when considering timelike geodesics in rotating black hole spacetime. In addition to the above-mentioned terms, time like geodesics should contain an additional term that should correspond to Coriolis force and most of these terms should depend on particle energy. 

In this paper, we investigate timelike geodesics in rotating Kerr and dirty spacetimes to find out how acceleration depends on energy and its sign. If we find that the sign of energy has an impact on geodesic motion, then one can distinguish particles with positive, negative, and zero energies through an experiment. Moreover, we consider the question about the existence of closed orbits for particles with zero energy in the ergoregion of a dirty rotating black hole. 

This work is organized as follows: in section 2, we briefly describe timelike geodesic motion in Schwarzschild spacetime, In sec. 3 we derive timelike geodesics for particles with zero energy in Kerr spacetime and find out their classical analogue. We prove that closed orbits for zero energy particles are absent in a general dirty rotating black hole in section 4. Finally, in section 5, we consider timelike geodesics in general axially-symmetric spacetime. Concluding remarks are displayed in section 6.

Throughout the paper, the geometrized system of units $G=c=1$ will be used. The signature is chosen to be $-,+,+,+$.

\section{Timelike geodesics in Schwarzschild spacetime}
Kerr black hole is a rotated version of a static spherically symmetric Schwarzschild black hole. For this reason, it is better to look deeper at geodesic motion in the Schwarzschild case to compare the obtained results in Kerr spacetime with the Schwarzschild black hole. The spherically symmetric Schwarzschild black hole is described by a line element
\begin{equation} \label{eq:metsch}
ds^2=-\left(1-\frac{2M}{r}\right)dt^2+\left(1-\frac{2M}{r}\right)^{-1}dr^2+r^2d\Omega^2,
\end{equation}
where $M$ is a mass of a black hole and $d\Omega^2=d\theta^2+\sin^2 \theta d\varphi^2$ is a metric on unit two-sphere. 
Static spherically-symmetric spacetime always admits two constants of motion i.e., energy-per-mass $E$ and angular momentum-per-mass $L$, which in Schwarzschild case in equatorial plane $\theta=\frac{\pi}{2}$ are given by
\begin{eqnarray} \label{eq:energysch}
-E&=&\left(1-\frac{2M}{r}\right)\frac{dt}{d\tau},\nonumber \\
L&=&r^2\frac{d\varphi}{d\tau}.
\end{eqnarray}
Here $\tau$ is the proper time of a particle. We consider timelike motion, thus  equations \eqref{eq:energysch} together with geodesic condition $g_{ik}u^iu^k=-1$ give the radial equation
\begin{eqnarray} \label{eq:potentialsch}
\left(\frac{dr}{d\tau}\right)^2+V_{eff}=0,\nonumber \\
V_{eff}=-\frac{2M}{r}+\frac{L^2}{r^2}-\frac{2ML^2}{r^3}+1-E^2.
\end{eqnarray}
Where $V_{eff}$ is an effective potential. We define forces acting on a particle as $-\frac{1}{2}\frac{dV_{eff}}{dr}$ and note the following points:
\begin{itemize}
\item The first term on right-hand-side of \eqref{eq:potentialsch} can be associated with Newtonian attractive force
\begin{equation}
a_N=-\frac{1}{2}\frac{d}{dr} \left(-\frac{2M}{r}\right)=-\frac{M}{r^2}.
\end{equation}
\item The second term in \eqref{eq:potentialsch} is repulsive centrifugal force
\begin{equation}
a_C=-\frac{1}{2}\frac{d}{dr}\left(\frac{L^2}{r^2}\right)=\frac{L^2}{r^3}.
\end{equation}
\item The third term represents general relativity corrections
\begin{equation}
a_{gr}=-\frac{1}{2}\frac{d}{dr}\left(-\frac{2ML^2}{r^3}\right)=-\frac{3ML^2}{r^4}.
\end{equation}
\end{itemize}
Thus, we need to find an effective potential in Kerr spacetime to compare it with Schwarzschild's effective potential \eqref{eq:potentialsch}.
\section{Forces for particles with zero energy in Kerr spacetime}    
An exterior  geometry of a rotating black hole is described by the Kerr metric which in Boyer-Lindquist coordinates has the following form:
\begin{equation}
	\begin{split}
		\label{metricKerr}
		&ds^{2}=-(1-\dfrac{2Mr}{\rho^{2}})dt^{2}-\frac{4Mra\sin^{2}\theta}{\rho^{2}}dt\,d\varphi+\frac{\rho^{2}}{\Delta}dr^{2}+\rho^{2}d\theta^{2}+ \\
		&+(r^{2}+a^{2}+\frac{2Mra^{2}\sin^{2}\theta}{\rho^{2}})\sin^{2}\theta d\varphi^{2},
	\end{split}
\end{equation}
Where
\begin{eqnarray} 		\label{constant}
\rho^{2}&=&r^{2}+a^{2}cos^{2}\theta,\nonumber \\
\Delta&=&r^{2}-2Mr+a^{2}
\end{eqnarray}
The Kerr metric is stationary and axially symmetric; it therefore admits the Killing vectors:
\begin{eqnarray}
-T^i&=&\frac{\partial x^i}{\partial t},\nonumber \\
\varphi^i&=&\frac{\partial x^i}{\partial \varphi} \,.
\end{eqnarray}
It is also asymptotically flat. The Komar formulae confirm that $M$ is the spacetime ADM mass, and show that $J=aM$is the angular momentum (so that $a$ is the ratio of angular momentum to mass).
In contrast to Schwarzschild spacetime, the equation $g_{00}=0$ doesn't define the event horizon location. This equation defines the surface where the Killing vector $\frac{d}{dt}$ becomes null. This surface is called an ergosphere or a static limit, and its location is $r_{e}=M+\sqrt{M^2-a^2\cos^2\theta}$. The Kerr spacetime possesses other two horizons i.e. the event $r=r_g$ and Cauchy horizons $r=r_C$,which locations are defined by $\Delta=0$ which yields:
\begin{eqnarray}
r_g&=&M+\sqrt{M^2-a^2},\nonumber \\
r_C&=&M-\sqrt{M^2-a^2} \,.
\end{eqnarray}
We don't consider here the extremal Kerr solution $M=a$ or naked singularity $M<a$, so, throughout the paper we assume that $M>a$. The region between the static limit and the event horizon is called an ergoregion. 

In the ergoregion, the Killing vector $\frac{d}{dt}$ is spacelike, however, the line element \eqref{metricKerr} is still timelike due to off-diagonal term $2g_{03}dtd\varphi$. This fact leads to the possibility that the particle has negative energy in this region. Penrose showed that particles with negative energy can extract energy from the rotational energy of a black hole~\cite{bib:pen}. One should realize that the energy $E$ is the particle energy regard to infinity, an observer inside the ergoregion will measure only particles with positive energy. 

We want to investigate the question about forces which act on a particle with zero and negative energy. For this purpose, we need to write down the geodesic equation. As we have mentioned above, the Kerr spacetime \eqref{metricKerr} admits two Killing vectors that lead to two constants of motion i.e. energy-per-mass $E$ and angular momentum-per-mass $L$ which can be obtained from the lagrangian
    \begin{eqnarray}
    		\label{energi1}
-p_{t}&=&\frac{\partial{\cal L}}{\partial{\dot{t}}}=g_{00}\dot{t}+g_{03}\dot{\varphi}=E=const, \\
    		\label{angularmoment1}
p_{\varphi}&=&\frac{\partial{\cal L}}{\partial{\dot{\varphi}}}=g_{33}\dot{\varphi}+g_{03}\dot{t}=L=const,
    \end{eqnarray}
    where $\cal L$ --- a lagrangian, and dot denotes a derivative concerning the proper time $\tau$. For simplicity, we restrict ourselves by consideration of the motion in the equatorial plane $\theta=\frac{\pi}{2}$. By using \eqref{energi1}, \eqref{angularmoment1} and \eqref{metricKerr}, one obtains the following first-order geodesic equations
    \begin{eqnarray}
    		\label{eq:vel2}
  \dot{t}&=&\dfrac{1}{\Delta}\left((r^{2}+a^{2}+2a^{2}M/r)E - 2MaL/r\right), \\
    		\label{eq:vel1}
\dot{\varphi}&=&\dfrac{1}{\Delta}\left(2MaE/r+L(1-2M/r)\right),
    \end{eqnarray}
  To find the radial component of the four-velocity, we use the geodesic condition $g_{ik}u^iu^k=-1$. Substituting into this condition \eqref{eq:vel2} and \eqref{eq:vel1} one obtains the equation
    \begin{equation}
    	\begin{split}
    		\label{velocityRpi/21}
    		\dot{r}^{2}=E^{2}+\dfrac{2M}{r}\left(aE-L\right)^{2}+\dfrac{a^{2}E^{2}}{r^{2}}-\frac{L^{2}}{r^{2}}-\dfrac{\Delta}{r^{2}},
    	\end{split}
    \end{equation}
    Introducing an effective energy $ 2E_{eff}=E^{2}-1$, we can write
    \begin{eqnarray}    		\label{conservationE}
\frac{\dot{r}^{2}}{2}&+&V_{eff}=E_{eff},\nonumber \\
V_{eff}&=&\frac{L^{2}}{2r^{2}}-\frac{M}{r}-\frac{M}{r^{3}}\left(aE-L\right)^{2}-\frac{a^{2}}{r^{2}}E_{eff}.
    \end{eqnarray}
    Comparing with the effective potential \eqref{eq:potentialsch} in the Schwarzschild case one can see the following points:
    \begin{itemize}
    	\item The leading term $-\frac{M}{r}$ corresponds to the Newtonian attractive force
    	\begin{equation}\label{eq:newtonianeq}
    		a_N=-\frac{M}{r^2} \,.
    	\end{equation}
    	This term is the same as in the Schwarzschild case. Note that for null geodesics this term is absent;
    	\item The term $\frac{L^2}{r^2}$ represents a repulsive centrifugal force
    	\begin{equation} \label{eq:centrifugaleq}
    		a_C=\frac{L^2}{r^3} \,.
    	\end{equation}
    	Which is again like in Schwarzschild's case;
    	\item The third term $-\frac{M}{r^3}\left(aE-L\right)^2$ is related to the relativistic correction of Einstein's general relativity, which accounts for the perihelion precession.
    	\begin{equation} \label{eq:greq}
    		a_{gr}=-\frac{3M}{r^4}\left(aE-L\right)^2 \,.
    	\end{equation}
    	Which has differences from the Schwarzschild black hole. We can consider the following cases:
    	\begin{itemize}
    		\item The first thing one can realize is that in the case of zero energy $E=0$ the term \eqref{eq:greq} coincides with Schwarzschild relativistic correction;
    		\item This term with negative energy $E$ and angular momentum $L$ is identical to the case of positive $E$ and $L$;
    		\item The maximal value it has in the case of positive energy $E>0$ and negative angular momentum $L<0$. Note that the case of negative energy $E<0$ and positive angular momentum $L$ is forbidden due to forward-in-time condition $\frac{dt}{d\tau}\geq 0$~\cite{bib:vertogradov2015gc}.
    	\end{itemize}
    	\item The term $-a^2E_{eff}/r^2$ depends on the black hole rotational parameter $a$ and absent in the Schwarzschild case. The acceleration
    	\begin{equation} \label{eq:koreoliseq}
    		a_{CF-}=\frac{2a^2}{r^2}E_{eff} \,,
    	\end{equation}
    	can be interpreted as the Coriolis force. Again, we can consider the following points:
    	\begin{itemize}
    		\item This force disappears for $E=\pm 1$;
    		\item it is negative when $-1 <E<+1$;
    		\item For null geodesic it is always positive except for the case of zero energy when it disappears. For null geodesics  \eqref{eq:koreoliseq} reduces to
    		\begin{equation}
    			a_{CF}^{null}=-\frac{2a^2E^2}{r^3} \,.
    		\end{equation}
    	\end{itemize}
    \end{itemize}

\section{Geodesics for particles with zero energy in general \- axially-symmetric spacetime}   
As we mentioned in the introduction the existence of circular orbits near the event horizon can lead to high-energy collisions. However, in Kerr spacetime, there are no closed orbits for particles with negative and zero energies in the ergoregion. For this reason, we decided to consider geodesic motion in general axially-symmetric black holes to find out if closed orbits for particles with zero energy can exist in the ergoregion. Thus, let us consider generally axially-symmetric spacetime which can be written, in equatorial plane $\theta=\pi/2$, in the form
     \begin{equation}
    	\begin{split}
    		\label{generalmetric}
    		ds^{2}=-N^{2}dt^{2}+R^{2}\left(d\varphi-\omega dt\right)^{2}+A^{-1}dr
    	\end{split}
    \end{equation}
Metric coefficients do not depend upon time $t$ and angle $\varphi$. As a result, this spacetime admits two Killing vectors related to time-translation and rotational invariance. It means that there are two constants of motion - Energy-per-mass $E$ and angular momentum-per-mass $L$. The event horizon location is defined as $N=0$\footnote{Note that there might be several roots of the algebraic equation $N=0$. However, we consider particle motion outside the event horizon. Thus, only the biggest root of $N=0$ is of interest.}
Metric components in Kerr spacetime have the following connection to general metrics in the form
    \begin{eqnarray}     		\label{connection}
R^{2}\omega^{2}&-&N^2=-1+\frac{2M}{r},\nonumber \\
\omega &=&\frac{2Mar}{(r^2+a^2)^2-a^2\Delta},\nonumber \\
R^{2}&=&r^2+a^2-\frac{2Ma^2}{r}.
    \end{eqnarray}
    in ergoregion $g_{00}>0$, then $N^{2}<\omega^{2} R^{2}$.\\
In what follows, all calculations will go close to notations introduced in~\cite{bib:on_negative}
      \begin{eqnarray}    		\label{generalgeo}
\dot{t}&=&\frac{X}{N^{2}},\nonumber \\
\dot{\varphi}&=&\frac{L}{R^{2}}+\frac{\omega X}{N^{2}},\nonumber \\
X&=&E-\omega L
    \end{eqnarray}
condition $u^{i}u_{i}=-1$  leads us to the radial geodesic equation
   \begin{eqnarray}    		\label{rgeneralgeo}
\frac{\dot{r}^{2}}{2}&+&V_{eff}=0,\nonumber \\
V_{eff}&=&-\frac{A}{N^{2}}Z^{2},\nonumber \\
Z^{2}&=&X^{2}-N^{2}-\frac{N^{2}}{R^{2}}L^{2},
    \end{eqnarray}
We are interested in circular orbits. In this case $\dot{r}=0$, which defines the  turning point $r=r_{tp}$. If additionally, $\ddot{r}\left(r_{tp}\right)=0$ then it defines the radius of a circular orbit, which we denote $r_c$
 \begin{eqnarray}     		\label{Condorbit}
  Z^{2}&=&0,\nonumber \\
(Z^{2})'&=&0.
    \end{eqnarray}
From zero energy ($E=0$) and from forward-in-time condition $\dot{t}\geq 0 \rightarrow X\geq 0$, one has
       \begin{eqnarray}    		\label{condE0}
X&=&-\omega L,\nonumber \\
\dot{t}&\geq &0 \rightarrow L\leq 0
    \end{eqnarray}
Condition for circular orbits then read
     \begin{eqnarray}     		\label{CondorbitE0}
 L_{c-}&=-&\frac{N}{\sqrt{\omega^{2}-\frac{N^{2}}{R^{2}}}},\nonumber \\
(Z^{2})'&=&\frac{N^{2}}{\omega^{2}-\frac{N^{2}}{R^{2}}}\left(\omega^{2}-\frac{N^{2}}{R^{2}}\right)'-2NN'.
    \end{eqnarray}
Let's introduce $\gamma=\omega^{2}R^{2}-N^{2}$ 
    \begin{eqnarray}
    		\label{1CondorbitE0}
(Z^{2})'&=&N^{2}\left(\ln{\left(\frac{\gamma}{N^{2}R^{2}}\right)}\right)',
    \end{eqnarray}
The condition $E=0$  leads to $\omega R>N$, then $\gamma>0$. 
The conditions $N'\geq0$, $R'\geq0$ follow from~\cite{bib:on_negative} and if $E=0$ then $\omega'<0$ which leads us to
       \begin{eqnarray}
    		\label{xicond1}
    \gamma'/\gamma<0 \rightarrow& (Z^{2})_{E=0}'<0,\\
    (\omega R)'<0 \rightarrow& \gamma'<0.
    \end{eqnarray}
Therefore, there are no circular orbits for particles with zero energy, in general, rotating dirty black hole spacetime. This result finalizes results obtained in ~\cite{bib:pavlov2015mpla, bib:vertogradov2015gc, bib:on_negative, bib:pavlov_zero} and can be formulated as the following statement: \textit{Closed orbits in the ergoregion of rotating black hole for particles with zero and negative energies are absent.}
\section{Forces in General Axially-Symmetric Spacetime}

In this section, we consider geodesic motion in general axially-symmetric spacetime. We assume that $A\sim N^{2}$, then effective potential   
\begin{equation}
    	\begin{split}
    		\label{force1}
    		V_{eff}=-(X^{2}-N^{2}&-\frac{N^{2}}{R^{2}}L^{2})
    	\end{split}
    \end{equation}
Like in Kerr spacetime, we define a fictitious force as $-\frac{1}{2}\frac{dV_{eff}}{dr}$ which yields a timelike geodesic equation in the form
\begin{eqnarray}
    		\label{force2}
\ddot{r}&=&-\frac{1}{2}V_{eff}',\nonumber \\
\ddot{r}&=&X\omega'L -N'N+L^{2}\left(\frac{NN'R-R'N^2}{R^3}\right)
    \end{eqnarray}
Without loss of generality, one can write $N^2$ in the form
\begin{equation}
N^2=1-\frac{2M(r)}{R}.
\end{equation}
Where $M(r)$ can be associated with the mass (or shape~\cite{bib:wisser}) function. In this case \eqref{force2} can be written as
\begin{eqnarray}\label{eq:force_final}
\ddot{r}&=&\left[\frac{1}{2}X\omega'L\right]_{I}-\left[\frac{MR'-M'R}{R^2}\right]_{II}\nonumber \\
&-&\left[\frac{L^2R'}{R^2}\right]_{III}+\left[\frac{3MR'-M'R}{R^4}\right]_{iv}.
\end{eqnarray}
Then we remark on the following points:
\begin{itemize}
\item The first term on the right-hand-side of \eqref{eq:force_final} can be associated with Coriolis force
\begin{equation}
a_{cor}=X\omega' L
\end{equation}
\item The second term represents a Newtonian attractive force
\begin{equation}
a_N=\frac{M'R-MR'}{R^2}.
\end{equation}
\item The third term is repulsive ($R'>0$) centrifugal force
\begin{equation}
a_C=\frac{L^2R'}{R^3}
\end{equation}
\item And final term on the right-hand-side of \eqref{eq:force_final} is general relativistic corrections
\begin{equation}
a_{gr}=\frac{3MR'-M'R}{R^4}.
\end{equation}
\end{itemize}
Thus, one can conclude that in general rotating dirty black hole spacetime, one has four groups of terms that define the particle motion, i.e.
\begin{itemize}
\item Newtonian attractive term;
\item Repulsive centrifugal term;
\item Coriolis term;
\item General relativistic corrections.
\end{itemize}
\section{Conclusion}

In this work, we explored different properties of geodesics for particles with zero energy in rotating Kerr and dirty black holes. If one considers null geodesics, then the radial motion in Kerr spacetime for zero energy particles coincides with radial motion in Schwarzschild spacetime. However, one can not consider geodesic motion in Schwarzschild spacetime as a mimic of geodesic motion in Kerr spacetime for zero energy particles because they have similarity only in the radial part. Other components of four-velocity do not coincide because they depend on energy and angular momentum. Considering timelike motion, we found that Newtonian attractive force and repulsive centrifugal force are the same as in the Schwarzschild case.
Moreover, it is not hard to show, that these terms are the same in general spherically-symmetric spacetime and its rotating version. General relativistic correction for zero energy particles and in Schwarzschild black hole are the same. These corrections might disappear when $aE=L$ for both negative and positive energies. However, this equality is forbidden by forward-in-time condition $\dot{t}\geq 0$. 
The distinctive feature of zero and negative energy particles is that they do not move along closed orbits in the ergoregion of Kerr black hole~\cite{bib:pavlov2015mpla, bib:vertogradov2015gc, bib:pavlov_zero}. The absence of closed orbits for particles with negative energies for generic rotating dirty black holes has been proven in the paper~\cite{bib:on_negative}. In this paper, we finalized this consideration by proving that zero energy particles do not move along closed orbits in the ergoregion of a generic rotating dirty black hole. That's why, one can make the following statement: \textit{circular and elliptical motion is forbidden for zero and negative energy particles.} Here, under the notion 'elliptical', we mean closed non-circular orbit.
The motion in the ergoregion has a lot of drawbacks. As we pointed out above zero and negative energies are just notions and an observer located in the ergoregion will always measure positive energy particles. However, the Penrose process allows to extraction of rotational energy from a black hole. Zero and negative energy particles can participate in high-energy processes in the horizon vicinity. All these processes might be observed through their influence on the visible size and specific intensity of a black hole shadow. To estimate the influence of zero and negative energy particles on specific intensity one needs to know the laws of their motion in the ergoregion of a rotating black hole. The geodesic properties obtained in this paper can help in exploring these observational features of a rotating black hole.

\textbf{acknowledgments}: This work is dedicated to the memory of our friend and teacher prof. Andrey A. Grib formulated this problem and gave us valuable comments.

\end{document}